\documentclass[conference]{IEEEtran}
\IEEEoverridecommandlockouts
\usepackage{cite}
\usepackage{amsmath,amssymb,amsfonts}
\usepackage{algorithmic}
\usepackage{graphicx}
\usepackage{textcomp}
\usepackage{xcolor}
\usepackage{multirow}
\usepackage{hyperref}
\usepackage{url}
\usepackage[caption=false,font=footnotesize]{subfig}
\def\BibTeX{{\rm B\kern-.05em{\sc i\kern-.025em b}\kern-.08em
    T\kern-.1667em\lower.7ex\hbox{E}\kern-.125emX}}
\begin{document}

\title{Evaluating Foveated Video Quality Using Entropic Differencing
\thanks{The authors thank Facebook Reality Labs for fruitful discussions and supports.}
}

\author{Yize Jin \\
\IEEEauthorblockA{
\textit{The University of Texas at Austin}\\
Austin, USA \\
yizejin@utexas.edu}
\and
\IEEEauthorblockN{Anjul Patney}
\IEEEauthorblockA{
\textit{Facebook Reality Labs}\\
Seattle, USA \\
rwebb@fb.com}
\and
\IEEEauthorblockN{Alan Bovik}
\IEEEauthorblockA{
\textit{The University of Texas at Austin}\\
Austin, USA \\
bovik@ece.utexas.edu}
}

\maketitle

\begin{abstract}
Virtual Reality is regaining attention due to recent advancements in hardware technology. Immersive images / videos are becoming widely adopted to carry omnidirectional visual information. However, due to the requirements for higher spatial and temporal resolution of real video data, immersive videos require significantly larger bandwidth consumption. To reduce stresses on bandwidth, foveated video compression is regaining popularity, whereby the space-variant spatial resolution of the retina is exploited. Towards advancing the progress of foveated video compression, we propose a full reference (FR) foveated image quality assessment algorithm, which we call foveated entropic differencing (FED), which employs the natural scene statistics of bandpass responses by applying differences of local entropies weighted by a foveation-based error sensitivity function. We evaluate the proposed algorithm by measuring the correlations of the predictions that FED makes against human judgements on the newly created 2D and 3D LIVE-FBT-FCVR databases for Virtual Reality (VR). The performance of the proposed algorithm yields state-of-the-art as compared with other existing full reference algorithms. Software for FED has been made available at: \url{http://live.ece.utexas.edu/research/Quality/FED.zip}
\end{abstract}


\section{Introduction}
In virtual reality (VR), immersive images and videos are important forms of media that carry visual information covering all $360$ degrees. Given the highest resolution of the human vision system (HVS) of about 120 pixels per degree (PPD), to reach a desirable viewing experience, the spatial resolution of immersive videos should reach at least 8K. In addition, immersive videos should include high frame rates (HFR) to reduce motion sickness, hence demanding even higher bandwidth consumption. 

To reduced the stresses imposed by such bandwidth-hungry immersive VR systems, foveated image / video compression is regaining relevance, due to the availability of consumer eye tracking devices that can be easily integrated into HMDs. Foveated image / video compression exploits the reduced visual acuity in the visual periphery, and assign higher compression ratio / quantization parameters (QP) to higher eccentricity relative to the fovea. Early implementations of the technique appeared more than two decades ago \cite{vpisc, fovmpeg, geisler1996, geisler1998}, however, neither the driving need nor affordable eye tracking devices were mature. 

Recently developed foveated compression algorithms, \cite{Illahi2020, Ryoo2016}, are usually supported by user studies which serve to verify the efficacy of the algorithms. However, although user / subjective evaluations are the most reliable method of quality assessment, they are also known to be expensive and time consuming, making it necessary to devise objective foveated quality assessment algorithms. Towards the development of objective algorithms, the authors of \cite{yizespie2020, yizetip2020} developed 2D and 3D foveated / compressed video quality databases in Virtual Reality, and evaluated a wide variety of quality assessment (QA) algorithms on the databases. Among available full reference (FR) foveated algorithms, Foveated Wavelet Quality Index (FWQI) \cite{fwqi} weights the error between the reference image and the foveated / distorted image in the wavelet domain, using a contrast sensitivity function (CSF) model \cite{geisler1998}, and a visually detectable noise threshold model \cite{watson1997}. The Foveation-based Content Adaptive SSIM (FA-SSIM) model \cite{fassim} weights the SSIM index with a foveation-based CSF which includes the retinal velocity and the corner velocity.

An important category of IQA algorithms relies on the statistical regularities of natural images (natural scene statistics -- NSS). NSS-based QA was first used in the Information Fidelity Criterion (IFC) \cite{ifc} and the Visual Information Fidelity (VIF) models \cite{vif}, where Gaussian scale mixtures (GSM) \cite{gsm} were used to characterize natural images in the wavelet domain, and image quality is captured by the conditional mutual information between the reference image model and the distorted image model. The creators of the Reduced Reference Entropic Differencing (RRED) \cite{rred} and Spatio-Temporal Entropic Differencing (STRRED) \cite{strred} models showed that, rather than computing conditional mutual information, which requires full reference information, a reduced amount of reference information can be obtained by computing the average differences of scaled local entropies in the wavelet domain. The authors of Speed-QA \cite{speedqa} showed that local entropic differencing can be applied in the spatial domain, achieving comparable performance to RRED or STRRED with much higher efficiency.

We propose an NSS-based foveated image quality assessment algorithm which model the bandpass-filtered images as GSM \cite{gsm}, and, for each subband, instead of scaling local entropies with local scalar factors, we weight the entropy map using a foveation-based error sensitivity function \cite{fwqi}, with frequency set to the average frequency of the subband. Then the entropy differences are calculated and averaged across subbands. We have found that the proposed Foveated Entropic Differencing (FED) algorithm achieves state-of-the-art performance on both the 2D and 3D LIVE-FBT-FCVR databases \cite{yizetip2020} when applied on a frame-by-frame basis.

\section{Foveation Based Error Sensitivity}
We employ the foveation-based error sensitivity function used in FWQI \cite{fwqi} for given frequency $f$ and eccentricity $e$:
\begin{equation}
    S_{f}(f,e) = \\
    \begin{cases}
        \frac{CS(f,e)}{CS(f,0)}=\exp(-0.0461f\cdot e), & \ f \leq f_{m}(e),\\
        0, & \ f > f_{m}(e),
    \end{cases}
    \label{eq:foverr}
\end{equation}
where $CS(f,e)$ is the contrast sensitivity function in \cite{geisler1998} defined by the inverse of the contrast threshold $CT$.
\begin{equation}
    CS(f,e) = \frac{1}{CT(f,e)}
\end{equation}
\begin{equation}
    CT(f,e) = CT_{0}\exp\left ( \alpha f \frac{e+e_{2}}{e_{2}}\right ),
    \label{eq:csf}
\end{equation}
where $CT$ is the contrast threshold, $CT_{0}$ is a constant minimal contrast threshold, $f$ and $e$ are spatial frequency and the eccentricity, respectively, $\alpha$ is the spatial frequency decay constant, and $e_{2}$ is the constant half-resolution eccentricity. We follow the parameter values given in \cite{geisler1998}: $CT_{0}=1/64, \alpha = 0.106$, and $e_{2}=2.3$. 

In (\ref{eq:foverr}) $f_{m}$ is the combined cutoff frequency for a given eccentricity:
\begin{equation}
    f_{m}(e) = \min(f_{c}(e), f_d)
\end{equation}
where the critical frequency $f_{c}$ for a given eccentricity $e$ is computed by setting $CT=1.0$ in (\ref{eq:csf}), which is the maximum possible contrast:
\begin{equation}
    f_{c}(e) = \frac{e_{2}\ln{(1/CT_{0})}}{(e+e_{2})\alpha}
    \label{eq:cutoff}
\end{equation}
and $f_d$ is the display Nyquist frequency: $f_d=d/2$ (cicles/degree), where $d$ is the display resolution:
\begin{equation}
    d \approx \frac{\pi Mv}{180} \;\text{(pixels/degree)}
\end{equation}
where $M$ is the with of the picture in pixel, $v$ is the viewing distance measured in image width, and can be related to the field of view (FOV) when gazing at the center of the image by:
\begin{equation}
    v = \frac{\cot(FOV/2)}{2}
\end{equation}

\section{The Proposed Algorithm}
NSS-based FR / RR IQA algorithms rely on the strong statistical regularities of natural images: bandpass coefficients of natural images follow heavy-tailed distributions \cite{gsm}, and divisively normalizing the bandpass coefficients using local variances causes the empirical distribution strongly tend towards a decorrelated normal Gaussian distribution \cite{ruderman}. The authors of \cite{reininger1983, rred, speedqa} have explored DCT decomposition, wavelet decompositions, and spatial predictive coding filters.

\begin{figure}[t]
    \centering
    \subfloat[Difference of Gaussian]{\label{fig:dog}\includegraphics[width=0.45\columnwidth]{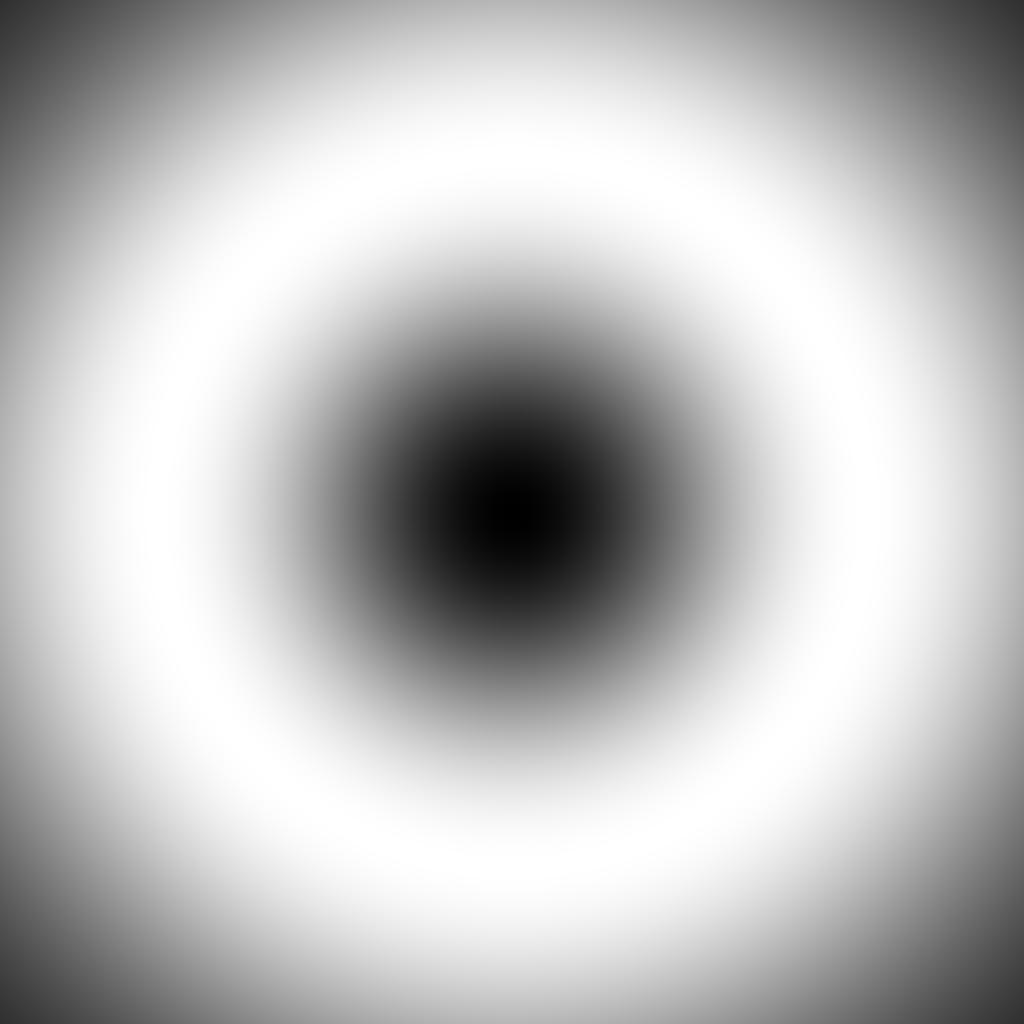}}\hfil
    \subfloat[Dirac-convolved Rectangular]{\label{fig:DcR}\includegraphics[width=0.45\columnwidth]{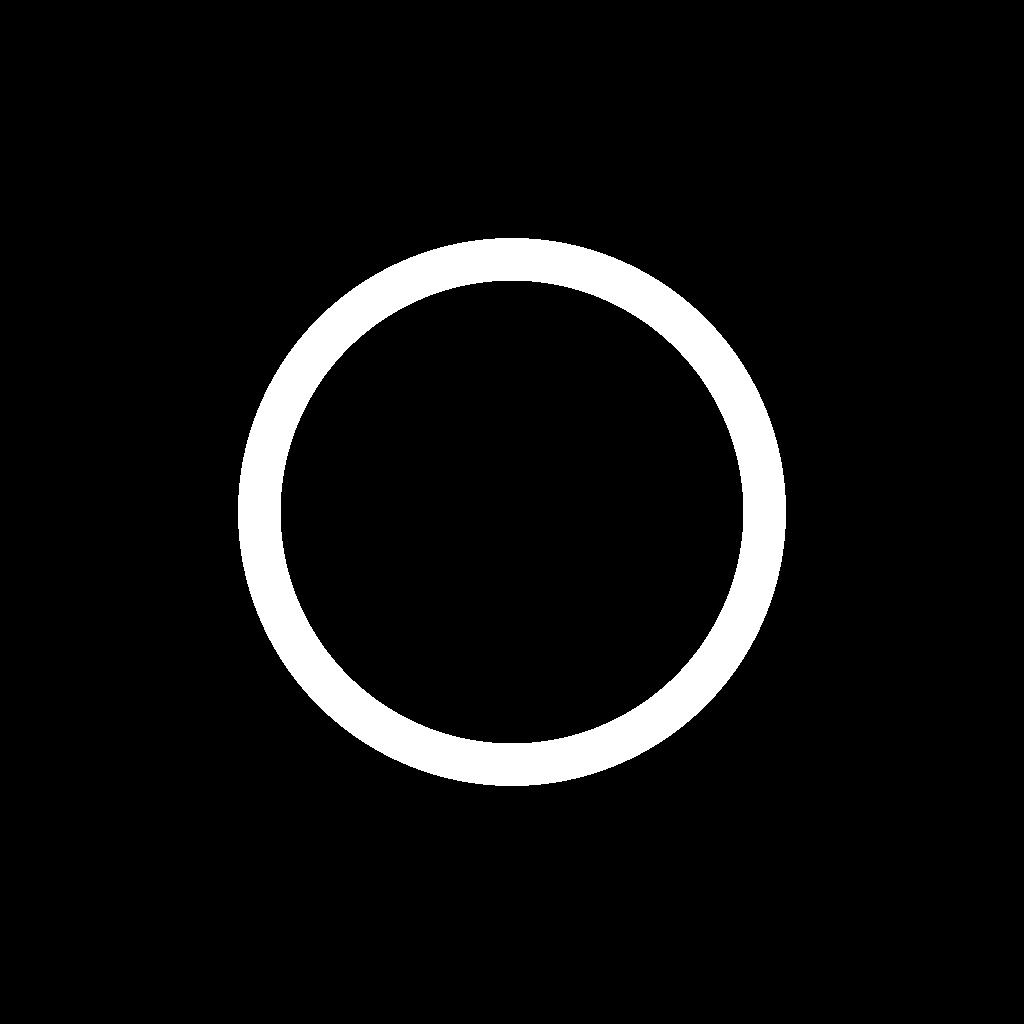}}\hfil
    \caption{A DoG and a 2D Dirac-convolved rectangular in the frequency domain. We set $\sigma_{1} = r_{1}$ and $\sigma_{2} = r_{2}$, where $\sigma_{1}$ and $\sigma_{2}$ are the two variances of Gaussians in the DoG, and $r_{1}$ and $r_{2}$ are the inner and outer radii of the Dirac-convolved rectangular. It can be seen that the DoG has much larger frequency coverage.}
    \label{fig:bandpass}
\end{figure}

The proposed FED model is based on the idea of combining measurements of statistical regularities (or lack thereof) with foveation-based error sensitivity functions. A wide range of zero-DC bandpass filters can be used to reveal these statistical regularities. We assume that when restricted to a narrow band in the frequency domain, as shown in Fig. \ref{fig:DcR}, the NSS of the bandpass coefficients often reflect distortions expressed in the particular frequency band. Therefore, the entropy differences computed from the NSS may be used to capture local perceptual distortions within subbands. A foveation-based error sensitivity function \cite{fwqi} tuned to a specific narrow subband can be applied to weight spatial entropy differences from each such subband to approximately weight the importance of perceptual distortions in the fovea and in the periphery.

\subsection{Modeling Bandpass Coefficients}
We define $\{B_{k}\}_{N}, k \in \{1,...,N\}$ to be a filterbank of zero-DC bandpass filters, and $\{\hat{I}_{k}\}$ corresponding bandpass reponses:
\begin{equation}
    \hat{I}_{k}(i,j) = B_{k}(i,j) * I(i,j),
    \label{eq:bandpass_conv}
\end{equation}
where $*$ and $I(i,j)$ are the convolution operator and the input image, respectively. We model a $b\times b$ bandpass response in $\hat{I}_{k}$ as a GSM vector: $\mathbf{x}\sim z\mathbf{U}$, where $\sim$ denotes equal in distribution, $z$ indicates the mixing multiplier which is independent of $\mathbf{U}$, and $\mathbf{U}$ is a zero-mean Gaussian vector with covariance matrix $\mathbf{C}_u$. The distribution of the GSM vector can be expressed as:
\begin{equation}
    f_{\mathbf{x}}(\mathbf{x}) \sim \int \frac{1}{(2\pi)^{d/2}|z^{2}\mathbf{C}_u|^{1/2}}\exp\left (\frac{-\mathbf{x}^{T}\mathbf{C_u}^{-1}\mathbf{x}}{z^{2}}\right )f_{z}(z)dz
\end{equation}

We found the choice of filterbanks also important. One possible choice is Difference of Gaussian (DoG) filters:
\begin{equation}
\begin{split}
    DoG&(r;\sigma_{1},\sigma_{2}) = \\ &\frac{1}{\sigma_{1}\sqrt{2\pi}}\exp\left(-\frac{r^{2}}{2\sigma_{1}^{2}}\right) - \frac{1}{\sigma_{2}\sqrt{2\pi}}\exp\left(-\frac{r^{2}}{2\sigma_{2}^{2}}\right),
\end{split}
\label{eq:dog}
\end{equation}
where $\sigma_{1}$ and $\sigma_{2}$ are standard deviations of the Gaussian functions, and $r$ is the distance between the pixel $(i,j)$ and the gaze center $(i_{0},j_{0})$:
\begin{equation}
    r = \sqrt{(i-i_{0})^{2}+(j-j_{0})^{2}}
    \label{eq:r}
\end{equation}
However, DoG filters are not desirable because of their large frequency coverage when trying to extract high-frequency information, as shown in Fig. \ref{fig:dog}. This is because when computing the CSF of the subband coefficients, it is desired that each bandpass filter only cover a narrow frequency band, so that the frequency band used in the CSF is precise. So we designed a bank of filters in the frequency domain by convolving an isotropic (circular) Dirac delta function with an ideal rectangular filter, as shown in Fig. \ref{fig:DcR}:
\begin{equation}
    \Tilde{B}_{k}(r) = \delta(r-r_{k}) * rect(r),
    \label{eq:freqfilt}
\end{equation}
where $\Tilde{B}_{k}$ is $B_{k}$ represented in the frequency domain, $r_{k}$ is the center frequency of the $k^{th}$ subband, $rect(r)$ is a rectangular function with width parameter $r_{b}$:
\begin{equation}
    rect(r) = \\
    \begin{cases}
        &1, \;|r|<r_{b},\\
        &0, \;\text{otherwise}
    \end{cases}
\end{equation}
The convolution in (\ref{eq:freqfilt}) can be easily computed as:
\begin{equation}
    \Tilde{B}_{k}(r) = \\
    \begin{cases}
        &1, \;|r-r_{k}|<r_{b},\\
        &0, \;\text{otherwise}
    \end{cases}
\end{equation}

\subsection{Foveated Entropic Differencing}
Let $C_{pqkr}$ and $C_{pqkd}$ be the $b\times b$ bandpass coefficients of the reference and distorted luminance image block $(p,q)$ using filter $B_{k}$ respectively. The bandpass coefficients are described by the GSM model:
\begin{equation}
    C_{pqkr} = Z_{pqkr}U_{pqkr}, C_{pqkd} = Z_{pqkd}U_{pqdk},
\end{equation}
where $Z_{pqkr}$ and $Z_{pqkd}$ are random scalar variables and $U_{pqkr}\sim \mathcal{N}(0,\mathbf{K}_{U_{kr}}$ and $U_{mkd}\sim \mathcal{N}(0,\mathbf{K}_{U_{kd}})$ are Gaussian random vectors. By passing the bandpass coefficients through a hypothetical neural noise model, we have:
\begin{equation}
    C_{pqkr}^{'}=C_{pqkr}+W_{pqkr},\;C_{pqkd}^{'}=C_{pqkd}+W_{pqkd},
\end{equation}
where $W_{pqkr}\sim \mathcal{N}(0,\sigma_{w}^{2}\mathbf{I}_{N})$ and $W_{pqkd}\sim \mathcal{N}(0,\sigma_{w}^{2}\mathbf{I}_{N})$. Since bandpass coefficients tends towards a multivariate Gaussian distribution when conditioned on realizations of $Z_{pqkr}$ and $Z_{pqkd}$, conditional local entropies are computed as:
\begin{equation}
    h(C_{pqkr}^{'}|z_{pqkr}) = const. + \frac{1}{2}\log[|z^{2}_{pqkr}\mathbf{K}_{U_{kr}}+\sigma_{w}^{2}\mathbf{I}_{N}|]
\end{equation}
\begin{equation}
    h(C_{pqkd}^{'}|z_{pqkd}) = const. + \frac{1}{2}\log[|z^{2}_{pqkd}\mathbf{K}_{U_{kd}}+\sigma_{w}^{2}\mathbf{I}_{N}|]
\end{equation}
In \cite{speedqa, rred} and \cite{strred}, the local entropies are scaled by scalar factors $\gamma_{pqkr}=\log(1+s_{pqkr}^{2})$ and $\gamma_{pqkd}=\log(1+s_{pqkd}^{2})$, here instead we weight the local entropies using the foveation based error sensitivity function. Specifically, let $f_{k}$ be the average frequency of the filter $B_{k}$, $e$ be the eccentricity in degrees of the block $(p,q)$, which is computed as follows:
\begin{equation}
    e(p,q) = \tan^{-1}\left(\frac{\sqrt{(bp-i_{0})^{2}+(bq-j_{0})^{2}}}{vM}\right)
\end{equation}
where $(i_{0}, j_{0})$ is the gazing center. We first normalize the foveation-base error sensitivity function (Equation (\ref{eq:foverr})) as:
\begin{equation}
    S_{k}^{\mathcal{N}}(p,q) = \frac{S_{f}(f_k,e(p,q))}{\sum_{p,q}S_{f}(f_k,e(p,q))}
\end{equation}
where we have assumed equal contribution of each subband. Then the foveated entropic differencing map for subband $k$ is defined as:
\begin{equation}
    D_{k}(p,q) = S_{k}^{\mathcal{N}}(p,q) \cdot (h(C_{pqkr}^{'}|z_{pqkr})-h(C_{pqkd}^{'}|z_{pqkd})),
\end{equation}
and the final quality prediction score is computed by taking the summation of the absolute entropy differences:
\begin{equation}
    FED = \sum_{k,p,q}|D_{k}(p,q)|
\end{equation}
\section{Experiments}
\subsection{Evaluation Framework}
The algorithms were tested and compared on the LIVE-FBT-FCVR databases \cite{yizetip2020}, where subjective opinion scores (MOS) and difference MOS (DMOS) were collected for 180 foveated / compressed images and 10 reference (190 total) 8K immersive videos. 

The foveation distortions in the subjective study of \cite{yizetip2020} were created in real time. During the playback, the system reads three immersive (equirectangular) videos, each video uniformly pre-compressed with one of 5 QP levels, and creates the foveation distortion by combining three videos with descending quality order from the fovea to the periphery, given gazing data obtained from the eye tracker. 

To recover the foveation experience, we adopt a viewport-based assessment framework as also mentioned in \cite{yizespie2020}, where for each immersive video we sample 18 viewports videos, and set the resolution of each viewport to 1024x1024, and the FOV to be $90^{\circ}$. The viewports distribute uniformly on a sphere in terms of longitude and latitude.

\subsection{Parameter Settings}
The display resolution $d$ and the display Nyquist frequency $f_{d}$ can then be computed as:
\begin{equation}
    d \approx \frac{\pi Mv}{180} = 8.94 \;\text{(pixels/degree)},
\end{equation}
\begin{equation}
    f_{d} = 4.47 \;\text{(cycles/degree)}
\end{equation}
We then divided the frequency band from $d$ to $0$ uniformly into $n$ subbands, and set $n=12$. We set the block size $b$ to 4 and the noise level $\sigma_{w}$ to 0.1.

\subsection{Overall Performance}
We compared FED with several leading existing IQA algorithms on both the 2D and 3D LIVE-FBT-FCVR databases: PSNR \cite{ssim}, SSIM \cite{ssim}, MS-SSIM \cite{msssim}, VIF \cite{vif}, S-RRED \cite{strred}, SpeedIQA \cite{speedqa}, FSIM \cite{FSIM}, FWQI \cite{fwqi}, and FA-SSIM \cite{fassim}. For each immersive video, each algorithm was evaluated on every viewport video on a frame-by-frame basis, and the scores obtained on each viewport frame were then averaged across all frames and 18 viewport videos. 

Pearson's linear correlation coefficient (PLCC), Spearman's rank order correlation coefficient (SROCC), Kendall's rank order correlation coefficient (KROCC), and root mean square (RMSE) between the DMOS and algorithm predictions are used as the criteria for evaluating the performance of the algorithms. Before computing PLCC and RMSE, a four-parameter logistic non-linearity was employed \cite{nllogistic}:
\begin{equation}
    Q(x) = \beta_{2} + \frac{\beta_{1}-\beta_{2}}{1+\exp(-\frac{x-\beta_{3}}{|\beta_4|})}
    \label{eq:12}
\end{equation}

For foveated QA, one intuitive solution would be first apply a foveated filter to both reference and distorted images, and use traditional QA methods to evaluated the filtered image pairs. So we also included as a baseline algorithm the method of using SpEED-IQA on foveated reference-distorted image pairs. We used the same foveated filtering method as in \cite{bradley2014}.

The results are shown in TABLE \ref{tab:overall}. The proposed FED model outperformed existing FR models, foveated and non-foveated, by a large margin on both 2D and 3D databases. In addition, the FED model also outperformed the intuitive foveated filtering model (FovFilt), showing that it can effectively capture foveation distortions. 

\begin{table}
\caption{Performance of algorithms on 2D and 3D LIVE-FBT-FCVR databases. The best performing algorithm is boldfaced.}
\begin{center}
\begin{tabular}{|c|c|c|c|c|}
\hline
\multirow{3}{*}{\textbf{IQA Models}}&\multicolumn{4}{|c|}{\textbf{Performances}} \\
\cline{2-5} 
& \multicolumn{2}{|c|}{\textbf{The 2D database}}& \multicolumn{2}{|c|}{\textbf{The 3D database}}\\
\cline{2-5} 
& \textbf{PLCC$\uparrow$}& \textbf{SROCC$\uparrow$} & \textbf{PLCC$\uparrow$} & \textbf{SROCC$\uparrow$}\\
\hline
PSNR \cite{ssim} & 0.6941 & 0.6954 & 0.4379 & 0.4418 \\
SSIM \cite{ssim} & 0.7260 & 0.7191 & 0.4184 & 0.4429 \\
MS-SSIM \cite{msssim} & 0.7288 & 0.7243 & 0.5337 & 0.5531 \\
VIF \cite{vif} & 0.8102 & 0.8068 & 0.6536 & 0.6765 \\
S-RRED \cite{strred} & 0.7896 & 0.7885 & 0.5604 & 0.5744\\
Speed-IQA \cite{speedqa} & 0.7760 & 0.7866 & 0.5084 & 0.4959\\
FSIM \cite{FSIM} & 0.7712 & 0.7808 & 0.5904 & 0.6273\\
\hline
FWQI \cite{fwqi} & 0.7906 & 0.7848 & 0.8041 & 0.7841\\
FASSIM \cite{fassim} & 0.7573 & 0.7418 & 0.7549 & 0.7401\\
\hline
FovFilt$^{\mathrm{*}}$ & 0.8484 & 0.8358 & 0.6000 & 0.6224\\
FED$^{\dagger}$ & \textbf{0.8971} & \textbf{0.8954} & \textbf{0.8276} & \textbf{0.8364}\\
\hline
\multicolumn{4}{l}{$^{\mathrm{*}}$Foveated filtering is simply applied before SpEED-IQA.}\\
\multicolumn{4}{l}{$^{\mathrm{\dagger}}$The number of subband is set to 12.}
\end{tabular}
\label{tab:overall}
\end{center}
\end{table}

\subsection{Number of Subbands}
We also studied the effects of using different numbers ($n=6,8,10,12$) of subbands, as shown in TABLE \ref{tab:subbandnum}. It can be seen that the performance is improved when using more subbands, since frequency information is more precisely represented, and foveation distortions are better captured. We also found that increasing the subband number over 12 leads to limited performance gain but higher complexity.

\begin{table}
\caption{Performance when using different numbers of subbands on the 2D LIVE-FBT-FCVR database. The best performing algorithm is boldfaced.}
\begin{center}
\begin{tabular}{|c|c|c|c|c|}
\hline
\multirow{2}{*}{\textbf{\#Subbands}}&\multicolumn{4}{|c|}{\textbf{Performances}} \\
\cline{2-5} 
& \textbf{PLCC$\uparrow$}& \textbf{SROCC$\uparrow$} & \textbf{KROCC$\uparrow$}$^{\mathrm{*}}$ & \textbf{RMSE$\downarrow$}\\
\hline
6 & 0.8701 & 0.8736 & 0.6900 & 4.73 \\
8 & 0.8860 & 0.8823 & 0.7030 & 4.57 \\
10 & {0.8930} & {0.8884} & {0.7109} & {4.43} \\
12 & \textbf{0.8971} & \textbf{0.8954} & \textbf{0.7196} & \textbf{4.35} \\
\hline
\multicolumn{4}{l}{$^{\mathrm{*}}$Kendall's rank order correlation coefficient.}
\end{tabular}
\label{tab:subbandnum}
\end{center}
\end{table}

\subsection{Using Different Filterbanks}
In addition to Dirac-convolved "ideal" low-pass filterbanks, we also tested DoG filterbanks, and a filterbank composed of isotropic Dirac-convolved triangular functions in the frequency domain. A rotationaly symmetric Dirac-convolved triangular frequency band introduces less Gibbs phenomena than using a rectangular frequency band, but also removes information in each subband. A Dirac-convolved triangular function is defined by:
\begin{equation}
    \Tilde{B}_{k}(r) = \\
    \begin{cases}
        &\frac{r_{b}-|r-r_{k}|}{r_{b}}, \;|r-r_{k}|<r_{b},\\
        &0, \;\text{otherwise}
    \end{cases}
\end{equation}
where $r$ is defined as in (\ref{eq:r}), and $r_{k}$ is the center frequency. Their performances are compared in TABLE \ref{tab:filterbanks}. It can be seen that, although there are severe spatial Gibbs phenomena when using Dirac-convolved rectangular functions, the statistics of the corresponding bandpass coefficients are less affected when capturing perceptual distortions. However, the results suggest that, for foveated quality assessment, large frequency bands (DoGs) and loss of information in each frequency band (isotropic Dirac-convolved triangular functions) may cause degraded performance.

\begin{table}
\caption{Performance when using different filterbanks on the 2D LIVE-FBT-FCVR database. The best performing algorithm is boldfaced.}
\begin{center}
\begin{tabular}{|c|c|c|c|c|}
\hline
\multirow{2}{*}{\textbf{Filterbanks}}&\multicolumn{4}{|c|}{\textbf{Performances}} \\
\cline{2-5} 
& \textbf{PLCC$\uparrow$}& \textbf{SROCC$\uparrow$} & \textbf{KROCC$\uparrow$}$^{\mathrm{*}}$ & \textbf{RMSE$\downarrow$}\\
\hline
DoG & 0.7768 & 0.7585 & 0.5720 & 6.18 \\
Triangular$^{\mathrm{\dagger}}$ & 0.8399 & 0.8302 & 0.6432 & 5.35 \\
Rectangular$^{\mathrm{\dagger}}$ & \textbf{0.8971} & \textbf{0.8954} & \textbf{0.7196} & \textbf{4.35} \\
\hline
\multicolumn{4}{l}{$^{\mathrm{*}}$Kendall's rank order correlation coefficient.}\\
\multicolumn{4}{l}{$^{\mathrm{\dagger}}$The function is convolved by n Dirac delta function.}
\end{tabular}
\label{tab:filterbanks}
\end{center}
\end{table}

\section{Conclusion}
We propose a foveated video quality model called FED that combines NSS-based QA features with a foveation-based error sensitivity function, based on the assumption that bandpass filtering restricts spatial entropy differences to capture perceptual distortions within a given frequency subband. We found that FED yields SOTA performance on the 2D and 3D LIVE-FBT-FCVR databases.

\bibliographystyle{IEEEbib.bst}
\bibliography{bibs/refs.bib}{}
\end{document}